\documentclass[a4paper,12pt]{article}
\pdfoutput=1 

\usepackage{jcappub} 
\usepackage[normalem]{ulem}
\usepackage{multirow}
\usepackage[utf8]{inputenc}
\usepackage{graphics}
\usepackage{epstopdf}
\usepackage{hyperref}
\usepackage{amsmath}
\usepackage{mathrsfs} 
\usepackage{bigints}
\usepackage{relsize}
\usepackage{psfrag}
\usepackage{epsfig}
\usepackage{subfig}
\usepackage{xcolor}
\usepackage{url}

\usepackage[T1]{fontenc} 
\usepackage[percent]{overpic}
\usepackage{breqn}

 \normalsize
\newcommand{\bmat}{\left(\begin{array}}
\newcommand{\emat}{\end{array}\right)}
\newcommand{\be}{\begin{equation}}
\newcommand{\ee}{\end{equation}}
\newcommand{\bea}{\begin{eqnarray}}
\newcommand{\eea}{\end{eqnarray}}

\usepackage{booktabs}
\usepackage{siunitx}
\def\lsim{\raise0.3ex\hbox{$\;<$\kern-0.75em\raise-1.1ex\hbox{$\sim\;$}}}
\def\gsim{\raise0.3ex\hbox{$\;>$\kern-0.75em\raise-1.1ex\hbox{$\sim\;$}}}

\title{\boldmath Waterfall phase in supersymmetric hybrid inflation
}

\author[a]{Ahmad Moursy,}
\author[b]{Qaisar Shafi}

\affiliation[a]{Department of Basic Sciences, Faculty of Computers and Artificial Intelligence,  \\
Cairo University, Giza 12613, Egypt}
\affiliation[b]{Bartol Research Institute, Department of Physics and Astronomy, \\
	University of Delaware, Newark, DE 19716, USA}
\emailAdd{a.moursy@fci-cu.edu.eg}
\emailAdd{qshafi@udel.edu}
\abstract{We explore a class of realistic supersymmetric hybrid inflation models with a predicted scalar spectral index $n_s \approx 0.97-0.978$, which is in good agreement with the recent Atacama Cosmology Telescope (ACT) measurement. The waterfall field responsible for the gauge symmetry breaking in this scenario experiences some $e$-foldings during the inflationary epoch. The scalar perturbations associated with the waterfall field during this phase induce a stochastic gravitational wave spectrum that will be tested in the ongoing Pulsar Timing Array (PTA) measurements and in future experiments. In an $SU(5)$ setting an observable number density of the superheavy GUT monopole linked to the waterfall field can be realized.
}
\begin{document}
\maketitle
\flushbottom

\section{Introduction}
\label{sec:intro}
Supersymmetric hybrid (SUSY) inflation models \cite{Dvali:1994ms,Copeland:1994vg,Linde:1997sj} provide an attractive framework for realizing a realistic inflationary scenario linked with gauge symmetry breaking. The minimal model utilizes a renormalizable superpotential that respects global $U(1)_R$ symmetry ($R$-symmetry), and a canonical K\"ahler potential. After taking radiative corrections into account, the minimal model predicts a scalar spectral index $n_s \approx 0.98$ \cite{Dvali:1994ms}, which is in reasonably good agreement with the recent ACT (Atacama Cosmology Telescope) measurement~\cite{ACT:2025tim}. The inclusion of a soft supersymmetry breaking term \cite{Rehman:2009nq,Senoguz:2004vu,Dvali:1997uq,Buchmuller:2014epa}, which appears during inflation, helps bring $n_s$ within the $0.97-0.98$ range~\cite{Rehman:2025fja}, which seems favored by the ACT measurement.

An extension of the minimal model employs a non-minimal K\"ahler potential~\cite{Bastero-Gil:2006zpr,urRehman:2006hu} which, in conjunction with the radiative corrections, also brings down $n_s$ within the preferred $0.97-0.98$ range. Recent studies have explored various hybrid inflation scenarios in light of Planck and ACT measurements; see~\cite{Moursy:2020sit,Moursy:2021kst,Ahmad:2025mul,Ahmed:2025crx,Pallis:2025epn,Okada:2025lpl}.

Hybrid inflation waterfall (WF) dynamics has received a significant amount of attention in both SUSY and non-SUSY frameworks~\cite{Clesse:2010iz,Kodama:2011vs,Clesse:2012dw,Clesse:2015wea,Moursy:2024hll,Lazarides:2023rqf,Maji:2024cwv,Spanos:2021hpk,Afzal:2024xci,Afzal:2024hwj,Tada:2023pue,Tada:2023fvd,Tada:2024ckk,Braglia:2022phb}, where inflation continues for a limited number of $e$-foldings during the WF phase. It was shown that the curvature perturbations are enhanced at small scales beyond the reach of current CMB observations. These amplified perturbations re-enter the horizon during the radiation dominated era, leading to the formation of sizeable density fluctuations that can gravitationally collapse to form primordial black holes (PBHs)~\cite{Garcia-Bellido:1996mdl,Clesse:2015wea}. Moreover, this enhancement also acts as a source of scalar induced gravitational waves (SIGW)~\cite{Ananda:2006af,Baumann:2007zm,DeLuca:2020agl,Gross:2024wkl}. In addition, the WF phase transition can play a role in partially diluting topological defects such as monopoles and cosmic strings~\cite{Lazarides:2023rqf,Moursy:2024hll,Maji:2024cwv,Maji:2025yms,Maji:2025thf}.

In this paper we explore a class of supersymmetric hybrid inflation models that respect $R$-symmetry with two important features. The radiative corrections can be safely ignored, and the waterfall field actively participates in the inflationary scenario {with a significant effect of the linear SUSY breaking soft term}. The desired fluctuations for large scale structure formation are again provided by the inflation field. However, on smaller scales the scalar field components in the waterfall field, under appropriate conditions, induce curvature perturbations that give rise to a stochastic gravitational wave spectrum that is compatible with the recent PTA measurements~\cite{NANOGrav:2023gor,EPTA:2023fyk,Xu:2023wog}. The predicted spectrum should be testable in the higher frequency range in a number of proposed experiments such as LISA, ET and DECIGO~\cite{LISACosmologyWorkingGroup:2025vdz,Bartolo:2016ami,Mentasti:2020yyd,Sato_2017}. Within this framework of supersymmetric hybrid inflation, the waterfall field in $SU(5)$ can play a key role in diluting the primordial GUT magnetic monopole number density to observable levels. 

Our paper is structured as follows. In Section~\ref{sec:SHI}, we revisit the scenario of supersymmetric hybrid inflation with minimal superpotential and non-minimal K\"ahler potential, where both respect a global $U(1)_R$ symmetry ($R$-symmetry). In Section~\ref{sec:inf-obs}, we explore a new region of parameter space of $R$-symmetric SUSY hybrid inflation scenario, with the following features: (1) inflation occurs close to the critical point with an adequate flat potential, (2) radiative corrections can be safely neglected, (3) SUSY breaking scale plays a significant role in the inflationary dynamics and observables, and (4) inflation continues during the waterfall for a limited number of $e$-foldings, such that the observable scales leave the Hubble radius when the waterfall fields are fixed at the origin. Thus, we calculate the inflationary observables analytically from a single field effective potential. Section~\ref{sec:WFdyn} addresses the dynamics during the waterfall and calculation of the power spectrum around the critical point following Refs.~\cite{Kodama:2011vs,Clesse:2015wea,Tada:2023pue}. In Section~\ref{sec:SU5}, we discuss waterfall dynamics and implications for SUSY $SU(5)$. We investigate the stochastic effects of multi-waterfall fields close to the critical point, and present the scalar induced gravitational waves spectrum. We show that for specific SUSY breaking scale, the gravitational waves spectrum can explain the recently observed PTA results. We estimate the $SU(5)$ monopole number density following its dilution during the waterfall phase.

%

\section{ Supersymmetric hybrid inflation }
\label{sec:SHI}
We consider the most general renormalizable superpotential that is invariant under a GUT gauge group $G$ and respects the global $U(1)_R$ symmetry  \cite{Dvali:1994ms}, 
\bea\label{Eq:suppot_tot}
W &=&   \kappa S \left( \Phi \, \overline{\Phi }  -  M^2  \right) ,
\eea
where $S$ is a gauge singlet chiral superfield under $G$, whose scalar component is the inflaton, with R-charge $R[S]=1$. The supermultiplets $\Phi$ and $\overline{\Phi }$ are conjugate representations of $G$, with $R[\Phi \, \overline{\Phi }] = 0 $. The waterfall (WF) field is included among the scalar components of $\Phi$ and $\overline{\Phi }$, which break the gauge group $G \to H$. We will denote the scalar components of the superfields by the same symbols of the corresponding superfields. The  K\"ahler potential with a non-minimal form is given by~\cite{Bastero-Gil:2006zpr}
\bea\label{Eq:K1}
K & = &  |S|^2 + |\Phi|^2 + |\overline{\Phi }|^2 + \kappa_S \frac{|S|^4 }{4 M_{\rm Pl}^2} + \kappa_{\Phi} \frac{ |\Phi|^4 }{4 M_{\rm Pl}^2}
+ \kappa_{\overline{\Phi }} \frac{ |\overline{\Phi }|^4 }{4 M_{\rm Pl}^2} + \kappa_{S\Phi} \frac{ |S|^2 (|\Phi|^2) }{ M_{\rm Pl}^2} + \kappa_{S\overline{\Phi }} \frac{ |S|^2 (|\overline{\Phi }|^2) }{ M_{\rm Pl}^2} \nonumber\\
&&+ \kappa_{SS} \frac{|S|^6 }{6 M_{\rm Pl}^4} + \cdots . %
\eea
Here $M_{\rm Pl}= 2.42 \times 10^{18}$ GeV denotes the reduced Planck mass. The F-term scalar potential is given by
\bea\label{eq:SUGRApot}
V_F = e^K \left[D_i W \, K^{i\bar{j}}\, D_{\bar{j}}\overline{W}- 3 \dfrac{|W|^2}{M_{\rm Pl}^2} \right],
\eea
where $i,j$ run over $(S,\phi^a,\bar{\phi}^a)$ with $\phi^a \, (\bar{\phi}^a)$ being the scalar components of $\Phi \, (\overline{\Phi})$, $a=1 \cdots {\cal N}$ and ${\cal N}$ denotes the
dimensionality of the representation of the waterfall field. $K^{i\bar{j}}$ is the inverse of the K\"ahler metric $K_{i\bar{j}}= \dfrac{\partial K}{\partial Z^i \partial Z^{\bar{j}}}$, and $D_i$ is the K\"ahler derivative defined by $D_i=\dfrac{\partial}{\partial Z^i} + \dfrac{ 1 }{M_{\rm Pl}^2}\dfrac{\partial K}{\partial Z^i}$. We work along the D-flat direction $|\Phi|= |\overline{\Phi }| $. The SUSY global minimum is located at 
\bea
\langle S\rangle= 0 \,\,\& \,\,\, \langle \Phi \, \overline{\Phi } \rangle= M ,
\eea 
at which, $D_i W = W =0$. At the true minimum, $G$ is broken to $H$ via the non-zero vev along the $H$ invariant direction $\phi$, while other components vanish during inflation and at the global minimum.

Throughout the inflationary epoch, the waterfall field $\phi$ is frozen at zero for $|S|>|S_c|= M $, due the large
positive mass squared of $\phi$, and at this stage, the vacuum energy $\kappa^2 M^4$ dominates the universe. It is well known that the minimal hybrid inflation framework as described above, including supergravity corrections, is naturally free from the $\eta$-problem~\cite{Bastero-Gil:2006zpr}.
Since SUSY is broken along the inflation trajectory, $\phi_a=0$, by non-zero values of $F_S$-term, a non-vanishing 1-loop radiative correction to the effective scalar potential is given by \cite{Coleman:1973jx}

\begin{eqnarray}
\Delta {V}_{\mathrm{CW}} = \frac{1}{64 \pi^2} \,\,\mbox{Str}\,
\left[ \mathcal{M}^4 (S ) \left( \ln\frac{\mathcal{M}^2
(S ) }{Q^2}-\frac{3}{2}\right) \right] \,,
\end{eqnarray}
where $\mathcal{M}^2 (S )$ is the
field-dependent mass-squared matrix of $\phi$, and $Q$ is a renormalization 
scale that we choose such that the radiative corrections vanishes at the critical point $S=S_c$ \cite{Bastero-Gil:2006zpr,Lazarides:2015cda}.
Along the trajectory where $\phi_a=0$, the radiative correction has the form \cite{Dvali:1994ms}
\begin{eqnarray}
\Delta {V}_{\mathrm{CW}}&\simeq& \frac{(\kappa M)^4}{8
  \pi^2} {\cal N} F(x) \,,\label{eq:V1loop}\\
F(x) &=& \frac{1}{4}\left[ (x^4+1) \ln \frac{(x^4-1)}{x^4}+2 x^2
  \ln \frac{x^2+1}{x^2-1} + 2 \ln \frac{\kappa^2 M^2 x^2}{Q^2}-3 \right] \,,
\end{eqnarray}
where $x= |S|/M$. Therefore, the necessary slope and curvature of the inflaton potential are provided by the radiative correction Eq.~(\ref{eq:V1loop}), the supergravity corrections in Eq.~(\ref{eq:SUGRApot}), as well as  soft SUSY breaking terms that have the form \cite{Dvali:1994ms}
\bea\label{eq:softSUSYB}
\Delta V_{\rm Soft}\simeq a \, m_{3/2} \, \kappa \, M^3 \, x + M_S^2 \, M^2 \,  x^2 .
\eea
Here $m_{3/2}$ is the gravitino mass, $a$ and $ M_S$ are the coefficients of soft SUSY breaking linear and mass squared terms for $S$ respectively, with $a= 2|2-A|\cos[{\rm arg} S + {\rm arg}(2-A)]$. In order to realize our inflationary scenario, we set $a=+1$. 
The complete potential in terms of the canonically normalized fields $\sigma= \sqrt{2} S$ and $\psi= \sqrt{2} \phi$ has the form
\bea\label{eq:totalpot}
V &=& V_F + \Delta {V}_{\mathrm{CW}} + \Delta V_{\rm Soft} \nonumber \\
&\simeq & \Lambda^4 \left[ \left( 1- \dfrac{\psi^2}{v^2}\right)^2   + \dfrac{2\sigma^2 \psi^2}{v^2 \, \sigma_c^2} + \frac{ \kappa^2 {\cal N}}{8
  \pi^2}  F(\sigma) +  \, \dfrac{  m_{3/2} }{ \sqrt{2} \, \Lambda^2 } \,\sigma
+\left( \dfrac{M_S^2}{2\Lambda^4} - \dfrac{\kappa_S}{2 M_{\rm Pl}^2} \right) \sigma^2 \right. \nonumber \\
&& \left. + \dfrac{\gamma_S}{8 M_{\rm Pl}^4} \sigma^4 + \cdots \right] ,	 
\eea
where $v= \sqrt{2} M $, $\gamma_S =  1 - \dfrac{7 \kappa_S}{2} +2 \kappa_S^2 - 3 \kappa_{SS}$, $\Lambda^2= \kappa M^2 = \kappa v^2 / 2$, $\sigma_c = v/\sqrt{2}$, and ellipses refer to higher order supergravity corrections that are suppressed. Here we have used the $R$-symmetry in order to rotate $S$ to the real axis. The masses at the true minimum are given by
\bea
m_\sigma = m_\psi \simeq 2\kappa M
\eea

When  $S$ reaches $S_c$, the mass squared of the waterfall field $\phi$ becomes tachyonic, which triggers the waterfall transition. At this time topological defects form if the homotopy group  of the manifold of degenerate vacuum states containing the quotient space $G/H$ is non-trivial. Cosmic strings are formed if the first homotopy group $\pi_1(G/H) \neq I$. Monopoles are formed if the second homotopy group $\pi_2(G/H) \neq I$, while domain walls arise when $\pi_0(G/H) \neq I$. If inflation ends at the waterfall transition, the formation of superheavy topological defects will be in conflict with current observations. Therefore it is necessary to partially/totally dilute its density by a certain number of $e$-foldings. 

We adopt a minimal setup, in which the WF phase continues for a limited number of $e$-foldings. We include the possibility that defects' densities are diluted to an observable level that can be probed  in the future experiments. In our scenario, inflation starts in the valley where $\phi$ is frozen at the origin and $\sigma$ drive the inflation for a limited number of  $e$-foldings that are sufficient for the cosmological scales to exit the horizon. At the critical value $\sigma_c$, WF starts, then continues for a specific number of $e$-foldings before inflation ends. We are interested in the regime where $|\sigma -\sigma_c| \ll 1$ during inflation.

We provide an example of a GUT gauge group $G\equiv SU(5)$ that is broken to $H\equiv SU(3)_c \times SU(2)_L \times U(1)_Y$. At the time when $\sigma=\sigma_c$, GUT monopole forms with a superheavy mass about one order of magnitude larger than the GUT symmetry breaking scale, which is a problem if inflation ends at the start of the waterfall. However, we let inflation continuing  for a suitable number of $e$-foldings after the waterfall such that the monopole density is diluted to an observable level that can be probed in the future experiments.

\section{Inflationary observables}
\label{sec:inf-obs}
Along the valley $\psi=0$, inflation is driven effectively by the gauge singlet $\sigma$. 
The inflationary observables are calculated at the time when the pivot scale $k_{*}= 0.05$ Mpc$^{-1}$ crosses the horizon, using the effective inflation potential before the waterfall transition that has the form
\bea\label{eq:infpot}
V_{\rm eff} &\simeq& \Lambda^4 \left[ 1+ \frac{ \kappa^2 {\cal N}}{8  \pi^2}  F(x) - 
\kappa_S \left( \dfrac{M^2}{M_{\rm Pl}^2}  \right) x^2 + \dfrac{\gamma_S}{2} \left( \dfrac{M^4}{M_{\rm Pl}^4}  \right) x^4  
+  \,  \left(\dfrac{m_{3/2}}{\kappa M}\right) x  \right.
  \nonumber \\
&& \left. +  \left(\dfrac{ M_S^2 }{\kappa^2 M^2 }\right)  x^2  + \cdots \right] , \nonumber \\
&\simeq& \Lambda^4 \left[ 1 + a_0 + \dfrac{(\sigma -\sigma_c)}{\mu_1} - \dfrac{(\sigma -\sigma_c)^2}{\mu_2^2} + \dfrac{(\sigma -\sigma_c)^3}{\mu_3^3} + \dfrac{(\sigma -\sigma_c)^4}{\mu_4^4}  \right],
\eea
where we have ignored the loop correction term for the small $\kappa$ values that we employ, and expanded around $\sigma_c$, in the last line of Eq.~(\ref{eq:infpot}). The parameters $\mu_i$ are dimension mass and $a_0$ is a dimensionless constant whose expressions in terms of the original potential parameters are given as follows
\bea
\mu_1  &=&  \left(  \frac{\sigma_c^3 \, \gamma_S}{2 M_{\rm Pl}^4} + 
\sigma_c^2 \left(  \frac{M_S^2}{\Lambda^4} -\frac{\kappa_S}{M_{\rm Pl}^2}  \right) + 
\frac{m_{3/2}}{  \sqrt{2} \Lambda^2}
 \right)^{-1}  \,,\nonumber\\
\mu_2  &=&   \left[  \frac{3 \sigma_c^2 \, \gamma_s}{4 \, M_{\rm Pl}^4} + \frac{1}{2} \left( \frac{M_S^2}{\Lambda^4 } - \frac{\kappa_S}{\, M_{\rm Pl}^2} \right)
 \right]^{-1/2}  \,,\nonumber\\
\mu_3  &=&   \left( \frac{ \gamma_S }{2 M_{\rm Pl}^4} \, \sigma_c\right)^{-1/3} \, ,  \nonumber \\
\mu_4  &=&     \left(\frac{ \gamma_S }{8 \, M_{\rm Pl}^4 }\right)^{-1/4}    \, ,  \nonumber  \\
a_0  &=& \frac{\sigma_c^4 \, \gamma_S}{8 M_{\rm Pl}^4} + 
\frac{1}{2} \sigma_c^2 \left( -\frac{\kappa_S}{M_{\rm Pl}^2} + \frac{M_S^2}{\Lambda^4} \right) + 
\frac{m_{3/2} \, \sigma_c}{ \sqrt{2 } \Lambda^2}.
\eea
In our analysis, the cubic and quartic terms in Eq.~(\ref{eq:infpot}), which are proportional to $\gamma_S$, have subdominant effects on the inflation dynamics and observables, since we choose $|\gamma_S| \lesssim 1$. With $|S|< M_{\rm Pl}$ we ensure that the SUGRA expansion is under control. The slow roll parameters are given by
\bea
\epsilon = \frac{M_{\rm Pl}^2}{2}\left( \frac{V_{\rm eff}'}{V_{\rm eff}}\right)^2  \, ,  \hspace*{1cm}
\eta = M_{\rm Pl}^2 \frac{ V_{\rm eff}''}{V_{\rm eff}} , 
\eea
where prime denotes the derivative with respect to $\sigma$. The observed amplitude of scalar perturbations is given by \cite{Planck:2018jri}
\bea
P_\zeta(k_*)= \frac{H^2_*}{8 \pi^2 \,  M_{\rm Pl}^2\, \epsilon_* } \simeq 2.1 \times 10^{-9} ,
\eea
where $H \simeq \dfrac{\Lambda^2}{\sqrt{3} M_{\rm Pl} } $ is the Hubble parameter during inflation. 

We assume that inflation starts close to $\sigma_c$ such that $|\sigma -\sigma_c|\ll 1$, and therefore $\epsilon \sim \dfrac{ M_{\rm Pl}^2}{2 \mu_1^2}$ and $\eta \sim -\dfrac{2 \, M_{\rm Pl}^2}{ \mu_2^2}$. In this case,
\bea
H &\simeq & 2.88\times 10^{-4} \,  \left( \dfrac{M_{\rm Pl}^2}{\mu_1 } \right), \\ 
\Lambda &\simeq & 0.0223 \, M_{\rm Pl} \left( \dfrac{ M_{\rm Pl} }{\mu_1} \right)^{1/2} .
\eea
 For a red tilted spectral index, a proper negative curvature of the potential, at the crossing horizon time, is provided by the coefficient of $\sigma^2$. It turns out that a spectral index $n_s$ consistent with the Planck~\cite{Planck:2018jri} and P-ACT-LB \cite{ACT:2025tim} results, is obtained with, $0.00475 < \dfrac{M_{\rm Pl}^2}{\mu_2^2} < 0.00955$. 
 
The amplitude of scalar perturbations is then given by
\bea\label{eq:PS1}
P_\zeta(k_*) & \simeq & \frac{\Lambda^4 \, \mu_1^2 }{12 \, \pi^2 \, M_{\rm Pl}^6 } , \nonumber \\
& = & \frac{\kappa ^2 }{ 6 \, \pi ^2 } \left( \frac{M}{M_{\rm Pl}} \right)^6 \left[ a \, \frac{ m_{3/2}  }{\kappa \,  M}   - 2 \,  \kappa_S \left( \frac{M}{M_{\rm Pl}}\right)^2  + 2 \, \gamma_S \left( \frac{M}{M_{\rm Pl}} \right)^4 \right]^{-2} \,,
\eea
and the tensor to scalar ratio $r$ and spectral index are given by
\bea\label{eq:r}
r = 16 \epsilon_* &\simeq & \dfrac{4 M_{\rm Pl}^2}{ \mu_1^2} = 4 \left[ a \, \frac{ m_{3/2} \, M_{\rm Pl} }{\kappa \,  M^2}   - 2 \,  \kappa_S \, \frac{M}{M_{\rm Pl} }  + 2 \, \gamma_S \left( \frac{M}{M_{\rm Pl}} \right)^2 \right]^{2} ,
\eea
\bea\label{eq:ns}
n_s = 1+2\eta_* - 6\epsilon_* &\simeq & 1- 4 \, \dfrac{ M_{\rm Pl}^2 }{\mu_2^2} = 1- 2 \kappa_S + 6 \gamma_S \, \left( \frac{M}{M_{\rm Pl}} \right)^2\, .
\eea
Here we have neglected the term with coefficient $M_S^2$ assuming that $ |{ M_S^2  }/({\kappa^2 M^2 } ) | \ll 1$. If the symmetry breaking scale $M  \, \lesssim {\cal O}(10^{16})$ GeV, the terms proportional to $\gamma_S$ are subdominant and can be neglected. Therefore, we obtain the relation\footnote{Equations~(\ref{eq:PS1}),~(\ref{eq:r}),~(\ref{eq:ns}),~(\ref{eq:PS2}) in terms of the original model parameters are derived independently and are in agreement with those in Ref.~\cite{Rehman:2017gkm}. However, the authors of Ref.~\cite{Rehman:2017gkm} discuss $\mu$-hybrid inflation that ends at the WF start.}
\bea\label{eq:PS2}
P_\zeta(k_*) & \simeq &  \frac{\kappa ^2 }{ 6 \, \pi ^2 } \left( \frac{M}{M_{\rm Pl}} \right)^6 \left[ a \, \frac{ m_{3/2}  }{\kappa \,  M}   - (1-n_s) \left( \frac{M}{M_{\rm Pl}} \right)^2 \right]^{-2} \,.
\eea

In Table~\ref{tab:obsevables} we provide three different benchmark points as well as the corresponding observables. The total number of $e$-foldings between the time of horizon crossing of the pivot scale and the time at the end of inflation, denoted by $\Delta N_*$, should be consistent with the thermal history of the universe. Assuming a standard thermal history, we have~\cite{Kolb:1990vq,Liddle:2003as,Ahmad:2025dds}
\begin{equation}\label{eq:Nethrmal}
\Delta N_* \simeq 53 + \frac{1}{3} \mathrm{ln} \left( \frac{T_r}{10^9 \, {\rm GeV}} \right) +\frac{2}{3} \mathrm{ln} \left( \frac{\Lambda}{10^{15} \, {\rm GeV} } \right),
\end{equation}
where $T_r$ is the reheating temperature.
%
\section{Dynamics during the waterfall phase transition}
\label{sec:WFdyn}
We are interested in the parameter space allowing inflation to continue after the start of the waterfall phase.\footnote{There are some studies in the literature of the dynamics after waterfall phase transition where primordial black holes and scalar induced gravitational waves may arise due to enhanced curvature perturbations~\cite{Clesse:2010iz,Kodama:2011vs,Clesse:2015wea,Moursy:2024hll,Lazarides:2023rqf,Maji:2024cwv,Spanos:2021hpk,Afzal:2024xci,Afzal:2024hwj,Tada:2023pue,Tada:2023fvd,Tada:2024ckk}.}
 The dynamics of $\sigma, \, \phi$ is determined by the potential
\bea\label{eq:totalpotexp}
\!\!\!\!\!\!\! V \simeq  \Lambda^4 \left[ \left( 1- \dfrac{\psi^2}{v^2}\right)^2   + \dfrac{2\sigma^2 \psi^2}{v^2 \, \sigma_c^2} + a_0 + \dfrac{(\sigma -\sigma_c)}{\mu_1} + \dfrac{(\sigma -\sigma_c)^2}{\mu_2^2} + \dfrac{(\sigma -\sigma_c)^3}{\mu_3^3} + \dfrac{(\sigma -\sigma_c)^4}{\mu_4^4} \right],
\eea
where we consider a single WF field $\psi$. We discuss the possible effects of the other scalar components of $\Phi$ and  $\overline{\Phi}$ in the following section.
The slow roll parameters during this stage are given by
\bea
\epsilon_i = \frac{ M_{\rm Pl}^2 }{2}\left( \frac{V_i }{V }\right)^2  \, ,  \hspace*{1cm}
\eta_{ij} = M_{\rm Pl}^2 \frac{ V_{ij} }{ V },
\eea
where $i,\,j$ runs over the fields $(\sigma, \psi)$. 

Close to the WF transition, $\psi $ is massless, and quantum fluctuations provide the initial displacement, $\psi_0 \sim H/2\pi$~\cite{Clesse:2010iz,Kodama:2011vs,Clesse:2015wea,Moursy:2024hll,Lazarides:2023rqf,Tada:2023pue}. We follow the analysis given in the latter references, where the integration of the quantum stochastic dynamics of the waterfall field $\psi$  is carried out using the slow-roll Langevin equation~\cite{Clesse:2015wea,Tada:2023pue} 
\bea
\partial_N \psi = - M_{\rm Pl}^2 \frac{V_\psi}{V} + \dfrac{H}{2 \pi} \xi_\psi(N),
\eea
where $\xi_\psi(N)$ is the independent noise with $\langle \xi_\psi(N) \xi_\psi(N^\prime)   \rangle = \delta(N-N^\prime)$.
A Gaussian distribution of the auxiliary field over a spatial region is considered, with
 the  width $\psi_0$ at the critical point of instability is  given by
\bea 
\psi_0^2 \equiv \langle \psi^2 \rangle = \dfrac{\Lambda^4 \, v \,\sqrt{\sigma_c\, \mu_1} }{48 \, (2\pi)^{3/2} M_{\rm Pl}^4 }.
\eea
It turns out that the SUSY breaking scale encoded in the gravitino mass and parameterized by $\mu_1$ 
plays a crucial role in the dynamics after the waterfall, as well as in determining the tensor to scalar ratio and the amplitude of scalar perturbations at the pivot scale. In order to study the dynamics during this stage, it is convenient to use the following field redefinitions 
\be 
\sigma \equiv \sigma_c \,  e^{\xi} \simeq \sigma_c (1+ \xi)  \,\,, \,\,\,\,\, \psi\equiv \psi_0  \, e^\chi .
\ee
Starting from the WF transition, the field dynamics encounter mainly two phases, Phases I and II~\cite{Clesse:2010iz,Kodama:2011vs,Clesse:2015wea,Tada:2023pue}. Considering the slow roll approximation after the waterfall, fields equations of motion can be written as 
 \bea 
 3 H\dot{\xi}\,  \sigma &=& -V_\sigma , \nonumber \\
  3 H\dot{\chi}\,  \psi &=& -V_\psi ,
 \eea
which can be recast in the form 
 \bea \label{eq:eom}
 {\xi}'  &=& - \dfrac{M_{\rm Pl}^2}{\sigma_c} \left[  \dfrac{1}{\mu_1} +  \dfrac{8 \sigma_c \psi_0^2 \, e^{2 \chi }}{v^4}\right] ,  \\
  {\chi}'  &=& - \dfrac{ 8 M_{\rm Pl}^2 }{v^2} \, \xi , 
 \eea
where the prime here denotes the derivative with respect to the number of $e$-foldings $N$, and we only consider leading terms in $\sigma - \sigma_c$. In phase I, the second term of the first equation of~(\ref{eq:eom}) is neglected, until grows and becomes dominant in phase II. The value of $\chi$ at the transition between phase I and II, denoted by $\chi_2$, is given by
\bea
\chi_2 =\ln\left( \dfrac{v \, \sqrt{ \sigma_c} }{ 2 \, \psi_0\, \sqrt{ \mu_1 }  } \right).
\eea
For phase I, the solutions are given by
\bea
\xi(N) &=& - \dfrac{M_{\rm Pl}^2}{\mu_1 \, \sigma_c} N  \,, \\
\chi(N) &=&  \dfrac{4 M_{\rm Pl}^4}{ v^2 \,\mu_1 \, \sigma_c} N^2 \,,  \\
\chi(\xi) &=&  \frac{4 \mu_1 \, \sigma_c}{v^2} \,  \xi^2  \,,
\eea
where we set $N=0$ at the time when $\sigma=\sigma_c$. Since $\sigma$ decreases, the value of $\xi$ at the end of Phase 2 has the form
\be 
\xi_2= - \frac{ v }{2 \sqrt{\mu_1 \, \sigma_c} } \, \chi_2^{1/2} \,.
\ee

Denoting by $N_1$, the number of $e$-foldings till the end of Phase I, we have
\be 
N_1 =  \dfrac{v \,  \, \sqrt{\mu_1 \, \sigma_c \, \chi_2} }{2 M_{\rm Pl}^2}.
\ee
\begin{figure}[htbp!]
    \centering
    \includegraphics[width=0.9\linewidth]{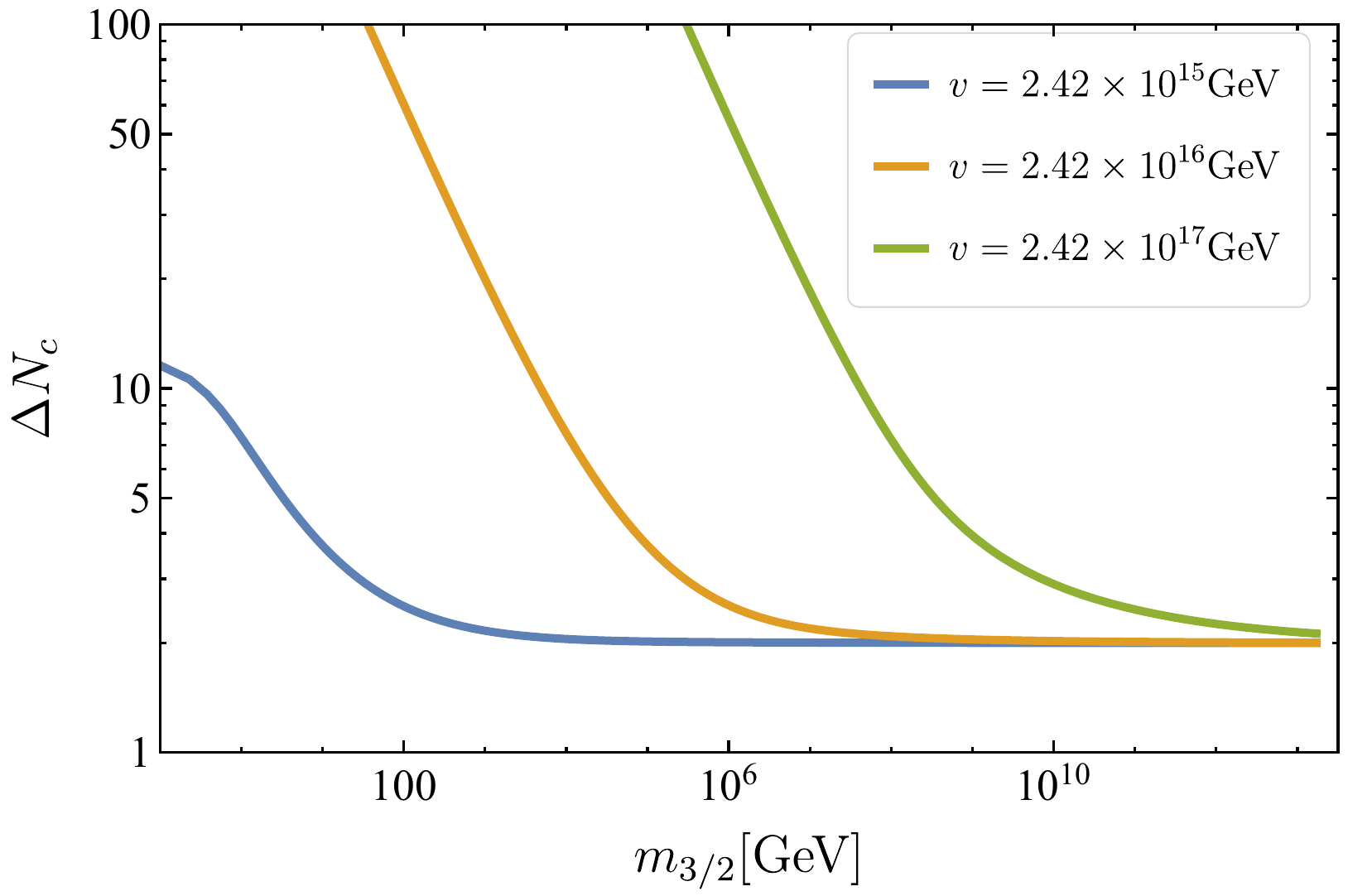}
    \includegraphics[width=0.9\linewidth]{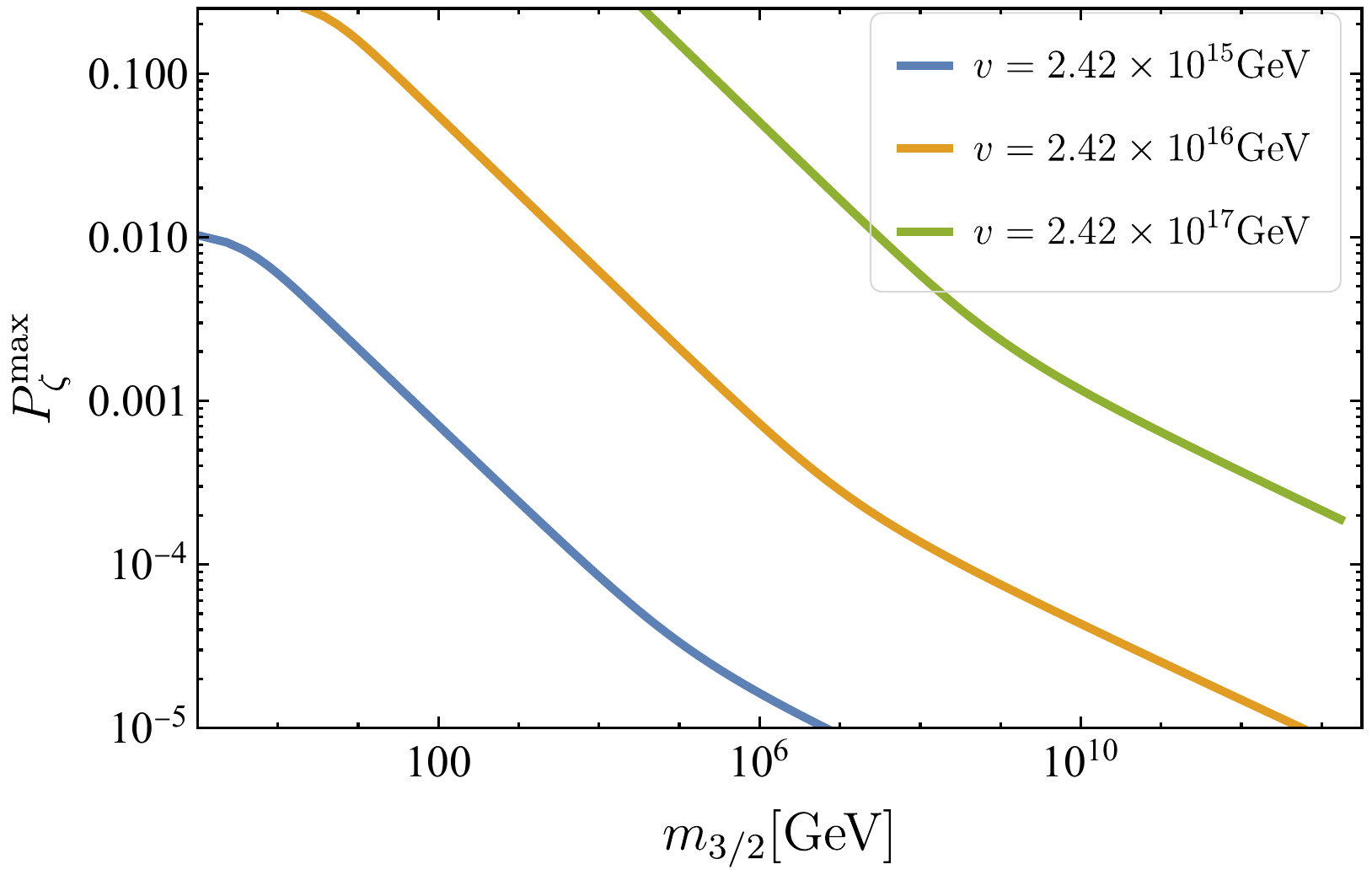}
      \caption{\label{fig:Nc-PR} Dependence on the gravitino mass $m_{3/2}$ of number of $e$-foldings that dilutes the monopole density during the WF phase, and peak amplitude of the power spectrum $P_\zeta^{\rm max}$ . We have set $n_s=0.973$ and vary the gauge symmetry breaking scales.}
\end{figure}
In phase II, the second term of the first equation of~(\ref{eq:eom}) is dominant and the solution is given by
\bea
\xi^2 = \xi_2^2 + \dfrac{v^2}{8 \mu_1 \, \sigma_c} \left[ e^{2(\chi - \chi_2 ) } -1  \right].
\eea
The number of $e$-foldings during the second phase is given by 
\bea 
N_2 =  \int_{\chi_{\rm end}}^{\chi_2} \, \dfrac{v^2}{8 \, M_{\rm Pl}^2 \, \xi} \, d\chi  
\simeq  \frac{v}{4 \, M_{\rm Pl}^2} \sqrt{\dfrac{\mu_1 \, \sigma_c }{\chi_2} } ,
\eea
where $\chi_{\rm end}$ is the value of $\chi$ at the end of inflation, determined at the time when $\eta_{\psi \psi} =1$. 

The total number of $e$-foldings during the WF till the end of inflation is $\Delta N_c = N_1 + N_2 $. It was shown that the number of $e$-foldings during the second phase is very small and can be neglected~\cite{Clesse:2010iz,Kodama:2011vs,Clesse:2015wea,Tada:2023pue}. It turns out from the above analysis that the the dynamics during the WF is sensitive to $\mu_1$ value (that controls $\Lambda$ and encodes the SUSY breaking), as well as the symmetry breaking scale $v$. 

In the $\delta N$ formalism \cite{Starobinsky:1985ibc,Salopek:1990jq,Sasaki:1995aw,Sasaki:1998ug,Lyth:2004gb,Sugiyama:2012tj}, the scalar power spectrum as a function of the mode wavelength $k$ is given by
\be
P_\zeta(k) = \dfrac{H^2_k}{4\pi^2} \left[ \left( \frac{\partial N_k }{ \partial \psi_k} \right)^2  +   \left( \dfrac{\partial N_k}{\partial \sigma_k} \right)^2   \right] ,
\ee
where $\psi_k \, (\sigma_k)$ are the values of $\psi \, (\sigma)$  when the mode with wavelength $k$ exits the horizon at $N=N_k$, where $k=a(N_k) H(N_k)$. 
In this case the power spectrum is given by \cite{Clesse:2015wea,Tada:2023pue,Spanos:2021hpk}
\be 
P_\zeta(k) \simeq \dfrac{\Lambda^4 \, v^2 \mu_1 \, \sigma_c}{192\,  \pi^2 \, M_{\rm Pl}^6 \, \chi_2 \, \psi_k^2     } \,,
\ee
with $\psi_k =\psi_0 \, e^{\chi_k}$, $\chi_k =\dfrac{4 \, M_{\rm Pl}^4 }{v^2 \, \mu_1 \, \sigma_c} (N_1+N_2 -N_k)^2 \,  $, and a peak  amplitude value, occurring at the point of instability, is given by
\be 
P_\zeta^{\rm max} \equiv  P_\zeta(k_c) \simeq \dfrac{\Lambda^4 \, v^2 \mu_1 \, \sigma_c}{192\,  \pi^2 \, M_{\rm Pl}^6 \, \chi_2 \, \psi_0^2     } \,.
\ee

Figure~\ref{fig:Nc-PR} shows the dependence on the gravitino mass of the number of $e$-foldings during the WF, and the peak amplitude of power spectrum.
In the following section, we provide an explicit example for realistic GUT monopoles, whose density is diluted by inflation continuing during the WF. In addition we study the accompanying scalar induced gravitational waves (SIGW).
%
\section{$SU(5)$ GUT monopoles and gravitational waves}
\label{sec:SU5}
In order to realized the importance of WF dynamics in realistic SUSY HI models, we consider an example of $SU(5)$ GUT model broken to the SM gauge group at the WF transition time, via the adjoint representation $\Phi \equiv \mathbf{24}_H$. The superpotential and K\"ahler potential are given by 
\bea\label{eq:suppotSU5}
W &=&   \kappa S \left( tr(\Phi^2)  -  M^2  \right) \nonumber \\ 
&=& \kappa S \left( \dfrac{1}{2}\sum_{a=1}^{24} \phi_a^2 -  M^2  \right) .
\eea
\bea\label{eq:K2}
K=  |S|^2 + tr|\Phi|^2 + \kappa_S \frac{|S|^4 }{4 M_{\rm Pl}^2} + \kappa_{\Phi} \frac{ (tr|\Phi|^2)^2 }{4 M_{\rm Pl}^2} + \kappa_{S\Phi} \frac{ |S|^2 (tr|\Phi|^2) }{ M_{\rm Pl}^2} + \kappa_{SS} \frac{|S|^6 }{6 M_{\rm Pl}^4} + \cdots , %
\eea
where $\Phi^\alpha_{\,\, \beta}\equiv \phi_a(T^a)^\alpha_{\, \beta}$, with the indices $a,b,c, \cdots=1,2,\cdots 24$, and $\alpha, \beta, \cdots=1,2,\cdots 5$, and $T^a$ are the $SU(5)$ generators. The critical value $\sigma_c = v/\sqrt{2}$, and the canonically normalized WF field $\psi \equiv \phi_{24}$ breaks $SU(5)$ to $SU(3)_c \times SU(2)_L\times U(1)_Y$. 

\subsection{Multifield stochastic effects}
\label{subsec:stochastic}
\begin{figure}[htbp!]
    \centering
    \includegraphics[width=0.95\linewidth]{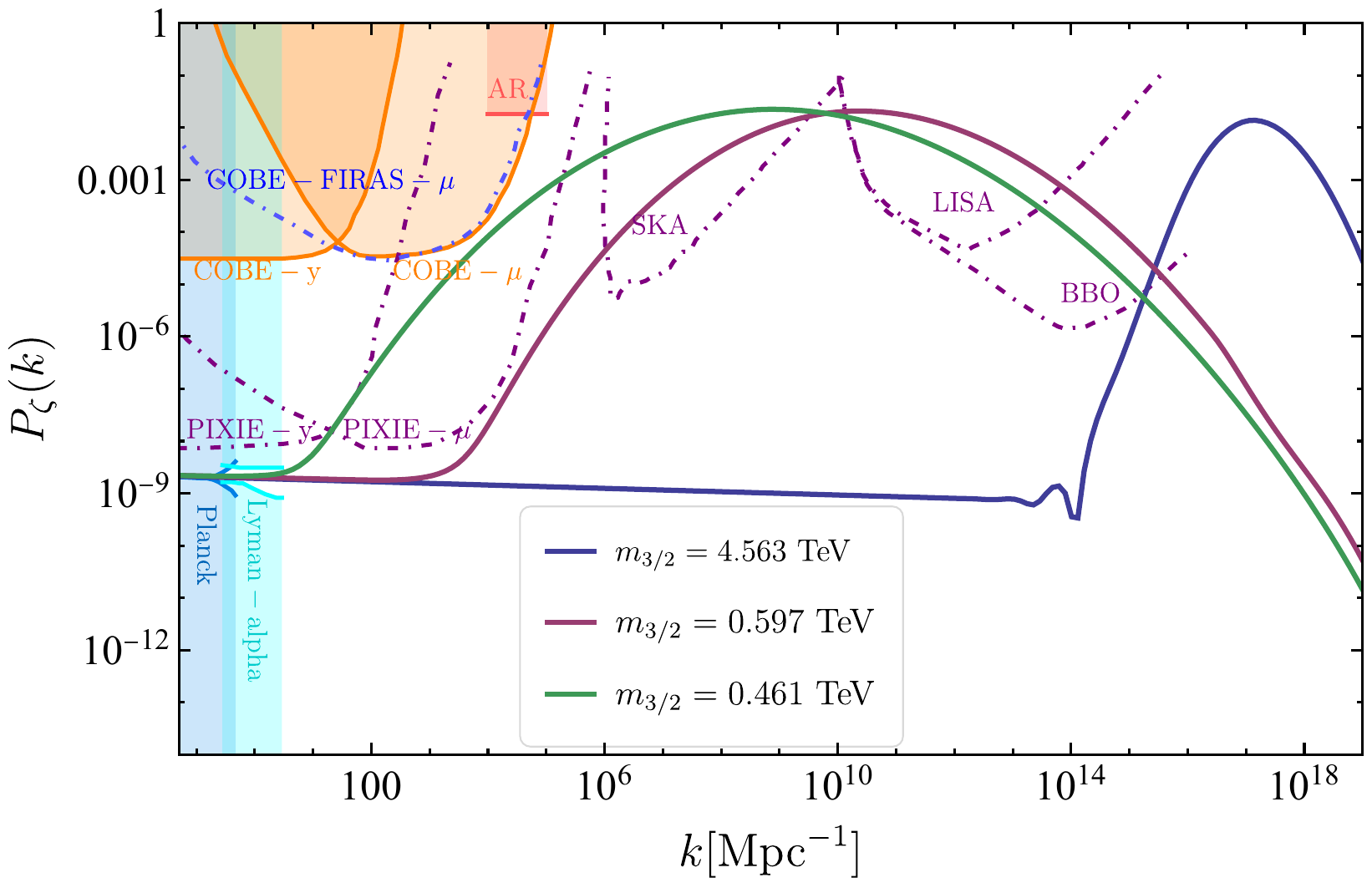}
        \caption{\label{fig:PS} Curvature power spectrum corresponding to benchmark points BP1, BP2 and BP3 with varying the gravitino mass and fixing the gauge symmetry breaking $v=2.42 \times 10^{16} $ GeV.}
\end{figure}
Around the critical point $\sigma_c$, WF fields (scalar components of $\Phi$ are massless and the potential is flat, therefore the dynamics of quantum fluctuations dominates over the classical background evolution, and stochastic effects of ${\mathscr{N}} $ real scalars should be taken into account~\cite{Clesse:2010iz,Clesse:2015wea,Kawasaki:2015ppx,Tada:2023pue,Tada:2023fvd,Tada:2024ckk}. At the time when WF starts, GUT symmetry is broken and the gauge fields corresponding to the broken generators acquire masses, with the longitudinal modes are equivalent to the massless scalars in the high-energy limit. If $g_{\rm GUT} \, \psi_r > H $, massive gauge fields are integrated out, where $g_{\rm GUT}$ is the unified gauge coupling value at the GUT scale, and $\psi_r $ is radial direction in the field space of WF fields defined by $\psi_r^2 \equiv \sum_i^{\mathscr{N}} \psi_i^2$. We adopt the assumption that the  calculations of curvature perturbations for the waterfall fields are using ${\mathscr{N}} $ multifield stochastic dynamics after the massive gauge bosons being integrated out for $e$-foldings around $N\sim N_c$~\cite{Tada:2024ckk}.

In our model the adjoint representation contains 24 complex degrees of freedom $ \phi^a $. Among these, the imaginary parts constitute 24 real degrees of freedom that acquire large masses squared $4 \Lambda^4\left( 2 v^2-  \psi_r^2 \right)/ {v^4}\gg H^2$, and can thus be integrated out. Furthermore, 12 real degrees of freedom are absorbed by the superheavy gauge bosons. Therefore, after $SU(5) $ symmetry breaking, at the waterfall start, the remaining number of real degrees of freedom is ${\mathscr{N}} = 12$ including $\psi \equiv {\rm Re}(\phi_{24})$. In this case the scalar potential possesses an $O({\mathscr{N}})$ symmetry and has the form

\bea\label{eq:totalpotSU5}
 V & \simeq &  
  \Lambda^4 \left[ 
  \left( 1-  \sum_{i=1}^{\mathscr{N}} \dfrac{\psi_i^2}{v^2}\right)^2   + \dfrac{2\sigma^2 }{v^2 \, \sigma_c^2} \sum_{i=1}^{\mathscr{N}}\psi_i^2 + a_0 + \dfrac{(\sigma -\sigma_c)}{\mu_1} + \dfrac{(\sigma -\sigma_c)^2}{\mu_2^2}  \right.
  \nonumber\\
  && \left. + \dfrac{(\sigma -\sigma_c)^3}{\mu_3^3} + \dfrac{(\sigma -\sigma_c)^4}{\mu_4^4} 
  \right],
\eea

A study of the stochastic dynamics of ${\mathscr{N}}$ waterfall fields based on the Fokker-Planck and Langevin equations has been done in Ref.~\cite{Tada:2023fvd,Tada:2024ckk}. It was shown that in the slow-roll limit, the dynamics reduces to a case with only the radial mode  and an effective centrifugal force proportional to ${\mathscr{N}}$. The Langevin equation for the radial direction in the slow-roll limit takes the form~\cite{Vennin:2015hra,Assadullahi:2016gkk,Tada:2023fvd,Tada:2024ckk}
\bea
    \partial_N \psi_r= -M_{\rm Pl}^2\frac{V_{\psi_r}}{V}+\frac{1}{2}  \left( \frac{H}{2 \pi} \right)^2  \frac{ {\mathscr{N}} -1}{\psi_r} +  \frac{H}{2 \pi}  \xi_{\psi_r}(N),
    \label{eq:langevin}
\eea
where $H^2 \simeq {V}/{3 M_{\rm Pl}^2}$, the normalized noise $ \langle \xi_{\psi_r}(N)\xi_{\psi_r}(N^\prime) \rangle =\delta(N-N^\prime)$, and the effective centrifugal force in the radial direction is represented by the second term on the right hand side that stems from the stochastic noise~\cite{Tada:2023fvd,Tada:2024ckk}. In this case, the amplitude of $\langle \psi_r^2 \rangle$ at the WF start time, denoted by $\psi_{r,c}^2$, is enhanced by a factor of ${\mathscr{N}}$, and is given by\footnote{The reader is referred to Refs.~\cite{Tada:2023fvd,Tada:2024ckk} for further details.}
\be
\psi_{r,c}^2 = \frac{ {\mathscr{N}} \Lambda^4 \Pi}{48 \sqrt{2 \pi^3} M_{\rm Pl}^2 },
\ee
with $\Pi \equiv \frac{v \sqrt{\mu_1 \sigma_c}}{M_{\rm Pl}^2}$. We use the value of $\psi_{r,c}$ as the initial condition at the start of WF when we solve the perturbations equations of motion. We follow the procedure in~\cite{Ringeval:2007am,Clesse:2013jra,Clesse:2015wea,Moursy:2024hll} for our numerical simulation of the perturbations dynamics described by the equations
\bea
	\delta \varphi_n^{''}+(3-\epsilon_H)\delta \varphi_n^{'}+\dfrac{1}{H^2} \sum_{m=1}^{3}V_{n m}\delta \varphi_m+\dfrac{k^2}{a^2H^2}\delta \varphi_n &= & 4\Phi_{\text{B}}^{'}\,\varphi'_n-\dfrac{2\,\Phi_{\text{B}}}{H^2}V_n, \\
	\Phi^{''}_{\text{B}}+(7-\epsilon_H)\,\Phi^{'}_{\text{B}}+\left(2\dfrac{V}{H^2}+\dfrac{k^2}{a^2H^2}\right)\Phi_{\text{B}}  & = &  - \dfrac{1}{H^2} \sum_{m=1}^{3} V_m\,\delta \varphi_m,
\eea 
where  $\varphi_n = (\sigma , \psi_r)$, $V_n$ is the derivative of $V$ with respect to $\varphi_n$, $k$ is the co-moving wave vector, and $\Phi_{\text{B}}$ denotes the Bardeen potential.\footnote{More details on the perturbations equations and calculation of the primordial power spectra can be found in Refs.~\cite{Ringeval:2007am,Clesse:2013jra,Moursy:2024hll,Maji:2024cwv}.} 
Here, we assume that during the classical dynamics $\psi_r$ aligns in the direction $\psi$ that breaks $SU(5)$ to the SM gauge group.

The scalar power spectrum $P_\zeta(k)$ is computed from the following formula \cite{Ringeval:2007am,Clesse:2013jra}
\bea
\label{eq:PR}
	P_\zeta(k)=\dfrac{k^3}{2\pi^2}\left|\Phi_{\rm B}+\dfrac{\sum\limits_{m=1}^{3} \varphi'_m\delta \varphi_m}{\sum\limits_{m=1}^{3}\varphi^{'2}_m}\right|^2.
\eea

We denote by $\Delta N_c$, the number of $e$-foldings between the onset of the WF transition and the end of inflation. To avoid gravitino overproduction, the reheating temperature must satisfy $T_r \lesssim 10^9$ GeV \cite{Ellis:1984eq,Moroi:1993mb,Kawasaki:2004yh}. We fix $T_r= 10^9$ GeV, set the soft SUSY breaking mass parameter $|M_S|\sim {\cal{O}}(1)$ TeV, and take the $SU(5)$ GUT symmetry breaking scale to be $v\simeq 2.42 \times 10^{16}$ GeV. The latter value of $v$ is consistent with the gauge coupling unification $SU(5)$~\cite{Ellis:1990wk,Amaldi:1991zx,Langacker:1991an,Amaldi:1991cn,Chakrabortty:2017mgi}.

In Figure~\ref{fig:PS}, we display the curvature power spectrum as a function of the wave number $k$ for various gravitino mass values represented by the three benchmark points in Table~\ref{tab:obsevables}. The constraints on the power spectrum from various current observations are incorporated, such as Planck~\cite{Planck:2018jri}, $\mu$- and $y$-distortions~\cite{Fixsen:1996nj}, and acoustic reheating (AR)\cite{Nakama:2014vla}. The dot-dashed curves represent the projected sensitivity of future PIXIE-like experiments to $\mu$- and $y$-distortions\cite{Kogut:2011xw}, along with the reach of upcoming gravitational wave observatories such as LISA, BBO, and SKA~\cite{Green:2020jor}.
%
\begin{table}[h!]
 \centering
 \begin{tabular}{| >{\centering\arraybackslash}p{3.0cm} |c | c | c | c | c |}
 \hline 
 \multicolumn{2}{|c|}{Parameters and observables}  & BP1 & BP2  & BP3 \\
 \hline 
 \multirow{6}{=}{{Potential parameters}} & $\Lambda^4[M_{\rm Pl}^4]$ & $1.38 \times 10^{-22}$ & $2.5 \times 10^{-24}$ &  $1.84 \times 10^{-24}$ \\
& $v$ [GeV] & $2.42\times 10^{16}$ & $2.42\times 10^{16}$  &  $2.42\times 10^{16}$\\
& $\kappa$ & $2.35 \times 10^{-7}$ & $3.51\times 10^{-8}$  & $2.71\times 10^{-8}$ \\
& $m_{3/2}$ [GeV] & $4.563 \times 10^{3}$ & $0.597 \times 10^{3}$ & $0.461 \times 10^{3}$ \\
& $\kappa_S$  & $ 0.016 $ & $0.014$ & $0.0141$ \\
& $\gamma_S$  & $0.944$ & $0.944$ &  $0.944$ \\
\hline\hline
\multirow{1}{=}{Masses in GeV} 
& $m_{\sigma} \, (m_\psi)$  & $5.68 \times 10^{11}$ & $8.49 \times 10^{10}$ & $6.56 \times 10^{10}$\\
\hline\hline
\multirow{5}{=}{Observables} & $A_s$ & $2.1\times 10^{-9}$ & $2.1\times 10^{-9}$ & $2.1\times 10^{-9}$ \\
& $n_s$ & $0.968$ & $0.972$ & $0.972$ \\
& $r$ & $4.44\times 10^{-16}$ & $9.67 \times 10^{-17}$ & $5.83 \times 10^{-17}$ \\
& $\sigma_*[M_{\rm Pl}]$ & $0.00707247$ & $0.00707118$ & 0.00707114\\
& $\Delta N_*$ & $49.8$ & $49.2$  & $49.1$ \\
& $\Delta N_c$ & $8.38$ & $23.5$  & $26.7$ \\
& $Y_M$ & $0.82\times 10^{-28}$ & $2.17\times 10^{-49}$  &  $1.01\times 10^{-53}$ \\
\hline 
 \end{tabular}
 \caption{Benchmark points with observables for $SU(5)$ GUT example with fixing the gauge symmetry breaking scale $v=2.42\times 10^{16}$ GeV. The reheating temperature is fixed at $T_r = 10^9 $ GeV, and the soft mass term parameter $|M_S| = 1$ TeV.}
 \label{tab:obsevables}
\end{table}
\newpage
\subsection{GUT Monopoles}
\label{subsec:monopole}
\begin{figure}[htbp!]
    \centering
      \includegraphics[width=0.85\linewidth]{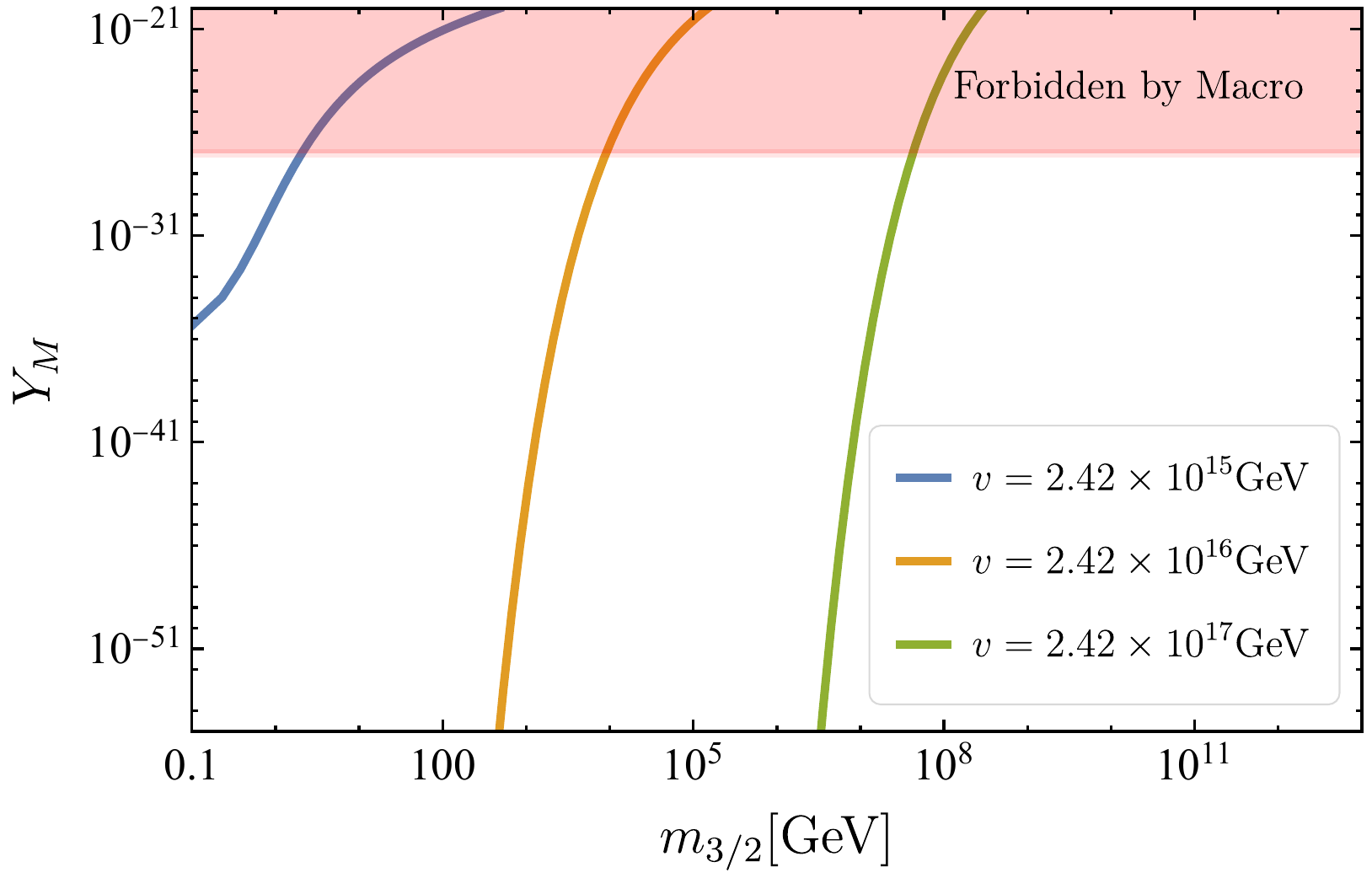}
    \caption{\label{fig:YM} Monopole yield $Y_M$ as a function of the gravitino mass $m_{3/2}$ for fixed $n_s=0.973$ and varying the gauge symmetry breaking scale.}
\end{figure}
At the start of the WF phase transition, the $SU(5)$ symmetry is broken and topologically stable monopoles are produced per unit correlation volume~\cite{tHooft:1974kcl,Polyakov:1974ek,Lazarides:1980cc,Shafi:1984wk}. The monopole that experiences a specific number of $e$-foldings, $\Delta N_c$, has a yield $Y_M$ after reheating that can be estimated from the following relation~\cite{Maji:2022jzu}   
\bea
Y_M = n_M/s\simeq \frac{45 \, \xi_G^{-3} \,  e^{-3 \, \Delta N_c} }{ 2\pi^2 \,  g_* \,  T_r^3}   \left( \frac{t_e}{t_r}\right)^2 ,
\eea 
where $n_M$ is the monopole number density, $s$ is the entropy density, $\xi_G \sim H^{-1}$ is the correlation length, $t_c$ is the time at the onset of the WF, $t_e$ denotes the time at the end of inflation, and $t_r$ is the time at reheating, which is related to the reheating temperature as follows~\cite{Lazarides:1997xr,Lazarides:2001zd}:
\be
T_r^2= \sqrt{\dfrac{45}{2\pi^2 \, g_*}} \, \dfrac{M_{\rm Pl}}{t_r},
\ee 
with $g_*= 206.25$  being the effective number of massless degrees of freedom at the end of reheating.

The monopole flux is constrained by upper bounds from several observations such as Parker bound~\cite{Parker:1970xv,Turner:1982ag,Adams:1993fj,Lewis:1999zm}, the MACRO experiment~\cite{Ambrosio:2002qq}, IceCube experiment~\cite{IceCube:2021eye}, and other observational constraints~\cite{Price:1983ax,Kajita:1985aig,Soudan-1:1986zji,Becker-Szendy:1994kqw,MACRO:2002iaq,Super-Kamiokande:2012tld,ANTARES:2022zbr}. We adopt the MACRO bound on the superheavy magnetic monopole of mass $m_M\gsim 10^{16} $ GeV~\cite{Ambrosio:2002qq}, namely the monopole present day flux $\Phi_M \lesssim 1.4\times 10^{-16}$ cm$^{-2}$ sec$^{-1}$ sr$^{-1}$. We can this into an upper bound on the monopole yield using the formula~\cite{Kolb:1990vq} 
\bea
Y_M \simeq 2 \times 10^{-10} \left( \frac{ \Phi_M }{{\rm cm }^{-2}\, {\rm sec}^{-1} \, {\rm sr}^{-1} } \right) \left( \dfrac{v_M}{10^{-3}} \right),
\eea
where $v_M$ is the monopole average speed. Therefore, the MACRO upper bound on the yield is $Y_M \lesssim 10^{-27}$.

It turns out that the monopole number density depends on the gravitino mass, which stems from the dependence of both the diluting $e$-foldings during the WF phase, as well as the correlation length dependence on the gravitino mass. In Figure~\ref{fig:YM}, we depict the dependence of the monopole yield on the the gravitino mass for various gauge symmetry breaking scales including the $SU(5)$ breaking scale of $2.42\times 10^{16}$ GeV. The light red region is forbidden by the observational bound from the MACRO experiment.
%
\subsection{Stochastic gravitational wave background}\label{subsec:SGWB}
\begin{figure}[htbp!]
    \centering
    \includegraphics[width=0.95\linewidth]{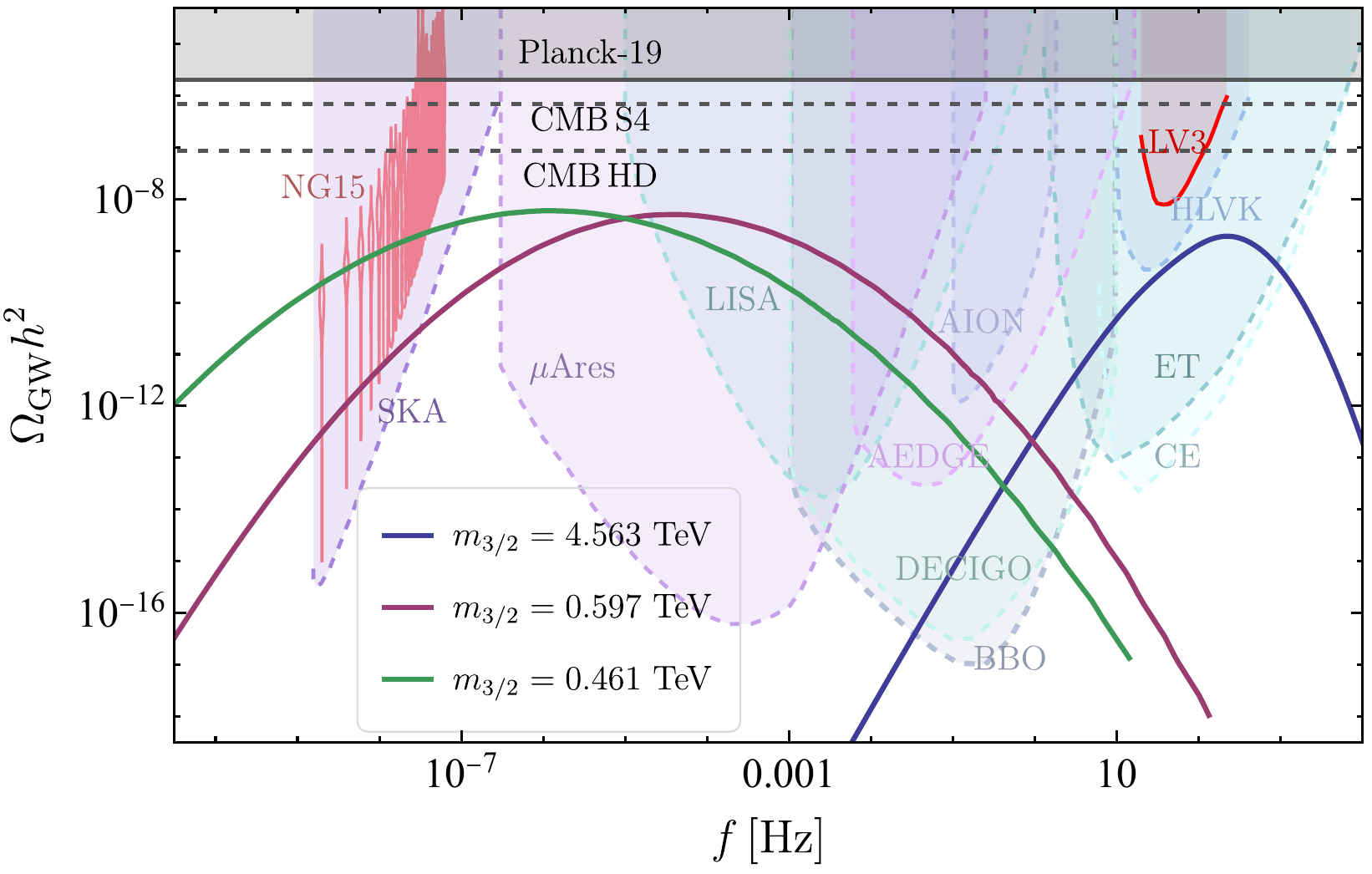}
    \caption{\label{fig:SIGW1} The scalar induced gravitational wave spectrum, in the SUSY $SU(5)$ model, as a function of frequency for various gravitino mass values. The symmetry breaking scale is set to $v=2.42\times 10^{16}$ GeV. The shaded regions depict the power-law integrated sensitivity curves for the ongoing and upcoming GWs detectors.}
\end{figure}
Scalar induced gravitational waves (SIGWs) arise as a second-order effect from enhanced curvature perturbations generated during the WF phase at small scales, well beyond the CMB observable window. This leads to the formation of sizeable density fluctuations upon horizon re-entry during the radiation dominated era. These enhanced scalar modes source tensor perturbations at second order in perturbation theory, generating a stochastic background of gravitational waves.\footnote{Primordial black holes can also be produced, but their abundance is highly suppressed in the case of SUSY $SU(5)$ model.}
 In terms of the primordial power spectrum, the scalar induced GWs spectrum is calculated via the formula \cite{Lewicki:2021xku,Kohri:2018awv,Espinosa:2018eve,Inomata:2019yww}
 \be \label{eq:OmegaGW}
\Omega_{\rm GW}^{\rm SI}h^2 \approx 4.6\times 10^{-4} \left(\frac{g_{*,s}^{4}g_{*}^{-3}}{100}\right)^{\!-\frac13} \!\int_{-1}^1 {\rm d} x \int_1^\infty {\rm d} y \, \mathcal{P}_\zeta\left(\frac{y-x}{2}k\right) \mathcal{P}_\zeta\left(\frac{x+y}{2}k\right) F(x,y) \bigg|_{k = 2\pi f} \,.
\ee
where 
\bea
F(x,y) &=&\frac{(x^2\!+\!y^2\!-\!6)^2(x^2-1)^2(y^2-1)^2}{(x-y)^8(x+y)^8} \times \nonumber \\
&&
\!\!\!\!\!\!\!\left\{\left[x^2-y^2+\frac{x^2\!+\!y^2\!-\!6}{2}\ln\left|\frac{y^2-3}{x^2-3}\right|\right]^{\!2} \!+\! \frac{\pi^2(x^2\!+\!y^2\!-\!6)^2}{4}\theta(y-\sqrt{3}) \right\}, 
\eea 
and we set $g_{*,s}\approx g_{*}$ in our calculations. 
\begin{figure}[htbp!]
    \centering
      \includegraphics[width=0.85\linewidth]{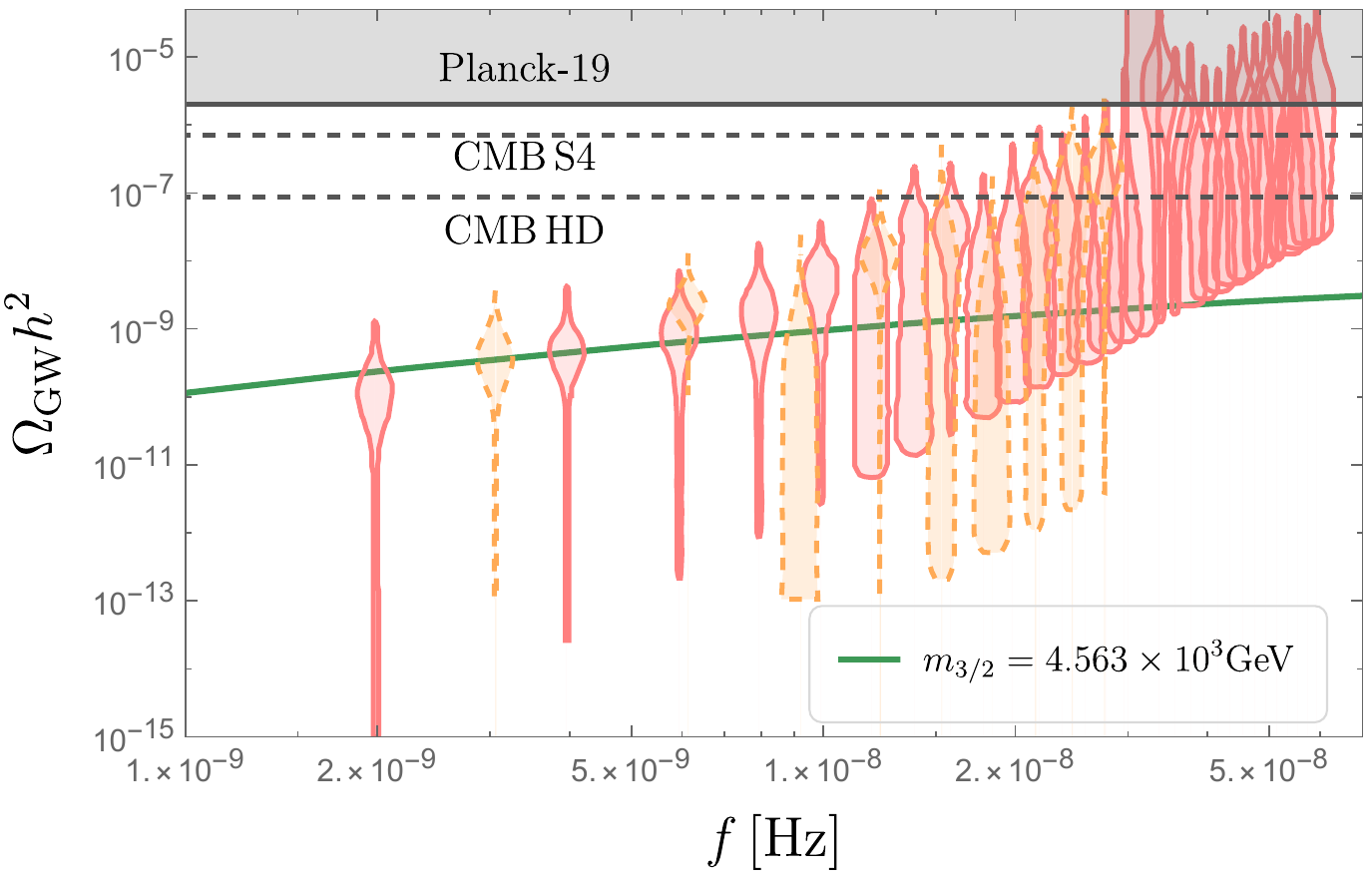}
    \caption{\label{fig:pta} Gravitational wave spectrum for gravitino mass $m_{3/2}=461$ GeV corresponding to BP3. Red violins represent the NANOGrav frequency bins, while the yellow ones represent EPTA frequency bins.}
\end{figure}

The resulting SIGW spectrum peaks at frequencies corresponding to the scale of the enhancement, which is correlated with the SUSY breaking scale encoded by the gravitino mass $m_{3/2}$. The SIGW spectrum for different values of $m_{3/2}$ fall within the sensitivity of future GW detectors, such as LISA \cite{Bartolo:2016ami, amaroseoane2017laser}, HLVK \cite{KAGRA:2013rdx}, ET \cite{Mentasti:2020yyd}, CE \cite{Regimbau:2016ike}, DECIGO \cite{Sato_2017}, AION~\cite{Badurina_2020}, BBO \cite{Crowder:2005nr, Corbin:2005ny}, AEDGE~\cite{Bertoldi:2019tck}, $\mu$-Ares~\cite{Sesana:2019vho} and SKA \cite{5136190, Janssen:2014dka}, as shown in Figure~\ref{fig:SIGW1}. Recent results from pulsar timing arrays such as NANOGrav \cite{NANOGrav:2023gor,NANOGrav:2023hvm} and EPTA \cite{EPTA:2023sfo,EPTA:2023fyk} are
shown as well. For $m_{3/2}=461$ GeV corresponding to BP3, the GWs spectrum can explain the PTAs results such as NANOGrav and EPTA, which is shown Figure~\ref{fig:pta}. However, the superheavy monopole density is extremely diluted with a flux $ \Phi_M \sim 10^{-40}$ cm$^{-2}$ s$^{-1}$ sr$^{-1}$. For BP2 with $m_{3/2}=597$ GeV, the GW spectrum lies in the LISA, DECIGO, and $\mu$Ares sensitivity regions, but the monopole density is very much diluted. For BP1, monopole density is partially diluted and can be observed in the next generation of monopole experiments, with the GW spectrum lying in the ET and CE sensitivity range.
%
\section{Conclusions}\label{sec:conc}
We have explored the important role the scalar waterfall field plays in a class of supersymmetric hybrid inflation models. The enhanced curvature perturbations produced during the waterfall phase induce a stochastic gravitational wave spectrum, which appears to be in good agreement with the measurements in the PTA frequency range. This prediction will be tested further in the higher frequency range in future experiments. In the framework of supersymmetric $SU(5)$, a limited number of $e$-foldings experienced by the waterfall field can dilute the number density of the primordial superheavy GUT monopole to observable levels.
%
\acknowledgments
We are grateful to Masaki Yamada for very helpful discussions, clarifications, and useful comments. The work of AM is partially supported by the Science, Technology \& Innovation Funding Authority (STDF) under grant number 50806.
%
\bibliographystyle{mystyle}
\bibliography{GUTs.bib}

\end{document}